\documentclass[aps,prd,a4paper,twocolumn,amsmath,showpacs,superscriptaddress,nofootinbib,preprintnumbers]{revtex4-1}

\usepackage{graphicx,ulem}
\usepackage{longtable}
\usepackage{float}
\usepackage{dcolumn}
\usepackage{graphics,epsfig}
\usepackage{amsmath,amssymb,latexsym,mathrsfs}
\usepackage{bm}
\usepackage{color}

\def\he4{$^4$He}
\def\h2{$^2$H}

\begin{document}

\title{Blue Gravity Waves from BICEP2 ?}


\author{Martina Gerbino}
\affiliation{Physics Department and INFN, Universit\`a di Roma 
	``La Sapienza'', Ple.\ Aldo Moro 2, 00185, Rome, Italy}

\author{Andrea Marchini}
\affiliation{Physics Department and INFN, Universit\`a di Roma 
	``La Sapienza'', Ple.\ Aldo Moro 2, 00185, Rome, Italy}

\author{Luca Pagano}
\affiliation{Physics Department and INFN, Universit\`a di Roma 
	``La Sapienza'', Ple.\ Aldo Moro 2, 00185, Rome, Italy}

\author{Laura Salvati}
\affiliation{Physics Department and INFN, Universit\`a di Roma 
	``La Sapienza'', Ple.\ Aldo Moro 2, 00185, Rome, Italy}

\author{Eleonora Di Valentino}
\affiliation{Physics Department and INFN, Universit\`a di Roma 
	``La Sapienza'', Ple.\ Aldo Moro 2, 00185, Rome, Italy}

\author{Alessandro Melchiorri}
\affiliation{Physics Department and INFN, Universit\`a di Roma 
	``La Sapienza'', Ple.\ Aldo Moro 2, 00185, Rome, Italy}

\date{\today}

\begin{abstract}
We present new constraints on the spectral index $n_T$ of tensor fluctuations from
the recent data obtained by the BICEP2 experiment.
We found that the BICEP2 data alone slightly prefers a positive, "blue", spectral index with
$n_T=1.36\pm0.83$ at $68 \%$ c.l.. However, when a TT prior on the tensor
amplitude coming from temperature anisotropy measurements is assumed we get
$n_T=1.67\pm0.53$ at $68 \%$ c.l., ruling out a scale invariant $n_T=0$ 
spectrum at more than three standard deviations.
These results are at odds with current bounds on the tensor spectral
index coming from pulsar timing, Big Bang Nucleosynthesis,
and direct measurements from the LIGO experiment.
Considering only the possibility of a "red", $n_T<0$ spectral
index we obtain the lower limit $n_T > -0.76$ at $68 \%$ c.l.
($n_T>-0.09$ when a TT prior is included).
\end{abstract}

\maketitle

\section{Introduction}\label{intro}

The recent detection of B-mode polarization made by the BICEP2 experiment
\cite{bicep2} clearly represents one of the major discovery in cosmology in the past
twenty years. While the BICEP2 result clearly needs to be confirmed by 
future experiments, it is timely and important to fully analyze the 
BICEP2 data and to identify all possible inconsistencies at 
the theoretical level.

In this brief note we focus our attention on the spectral index of
tensor fluctuations $n_T$.
Indeed, a crucial prediction of inflation is the production of a stochastic
background of gravity waves (\cite{GWs}) with a slightly tilted spectrum,

\begin{equation}
n_T = -2\epsilon ~,
\end{equation}

\noindent where $\epsilon=-\dot{H}/H^2$ denotes a slow roll 
parameter from inflation ($H$ is the Hubble rate during the inflationary stage).

In standard inflation $\epsilon$ is strictly positive~\cite{lr} and in the 
usual  parameter estimation routines, the tensor spectral index
is assumed to be ``red'', or negligible.

However, in recent years, a set of inflationary models has
been elaborated where the spectral index of tensor modes could be
positive, $n_T>0$, i.e. ``blue''.
A first attempt to compare these models with observational data has been
made in \cite{camerini}.

The main theoretical problem for the production of a blue spectrum of gravitational 
waves (BGW) is that the stress-energy tensor must violate the so-called Null Energy Condition
(NEC). In a spatially flat FRW metric, a violation of NEC indeed
corresponds to the inequality $\dot{H}<0$ and is ultimately the reason for the red 
tensor spectrum in standard inflation.

Models that violates NEC have been already presented. For example, in the so-called 
super-inflation models \cite{super} where inflation is driven by a component 
violating the NEC a BGW spectrum is expected. 
Models  based on  string gas cosmology as in \cite{stringgas}, where
scalar metric perturbations are thought to originate from initial string
thermodynamic fluctuations \cite{stringpert}, also can explain a BGW background.
A BGW spectrum is also a generic prediction of a class of four-dimensional
models with a bouncing phase of the universe \cite{bounce}. To induce the
bounce, the stress-energy tensor must violate the null energy condition
(NEC). G-inflation \cite{yoko}, has a Galileon-like nonlinear 
derivative interaction in the  Lagrangian with the resultant equations of 
motion being of second order. In this model, violation of the null energy condition 
can occur  and the spectral index of tensor modes can be blue. 
BGW may also be present in scalar-tensor theories and $f(R)$ gravity theories.

It is therefore timely to investigate the constraints on the tensor spectral 
index $n_T$ from the BICEP2 data. Strangely enough, no constraint on this
parameter has been presented by the BICEP2 collaboration while, as we discuss
in the next section, we found that the BICEP2 data could provide interesting
results on this parameter.

\section{Analysis method}\label{analysis}
Our analysis method is based on the Boltzmann CAMB
code~\cite{Lewis:1999bs} and a Monte Carlo Markov Chain (MCMC) analysis based
on the MCMC package \texttt{cosmomc}~\cite{Lewis:2002ah} (version December 2013).
We have implemented in the MCMC package the likelihood code provided
by the BICEP2 team (we just use BB data).
and considered as free parameters the
ratio of the tensor to scalar amplitude $r$ at $0.01 hMpc^{-1}$, 
defined as $r_{0.01}$, and the tensor spectral index $n_T$.
We prefer to use the pivot scale at $k=0.01 hMpc^{-1}$ since
the BICEP2 data is most sensitive to multipole $l \sim 150$
and using the approximate formula $l\sim 1.35\times10^4 k$.

All the remaining parameters have been kept fixed at the Planck+WP
best fit values for the LCDM+r scenario (see \cite{planck2013}).

Moreover, since the tensor amplitude should also be consistent with the
upper limits on $r$ coming from measurements of the temperature
power spectrum, we have assumed a prior of $r_{0.002}<0.11$ at $95 \%$ c.l. 
(see \cite{planck2013inf}). We refer to this prior as the
"TT" prior.

Note that the TT prior is taken at much larger scales, $k=0.002 hMpc^{-1}$ than
those sampled by the BICEP2 experiments. As we show in the next section 
this prior is extremely important for the constraints on $n_T$.

\section{Results} \label{results}

\begin{figure*}
\begin{tabular}{c c}
\includegraphics[width=6cm]{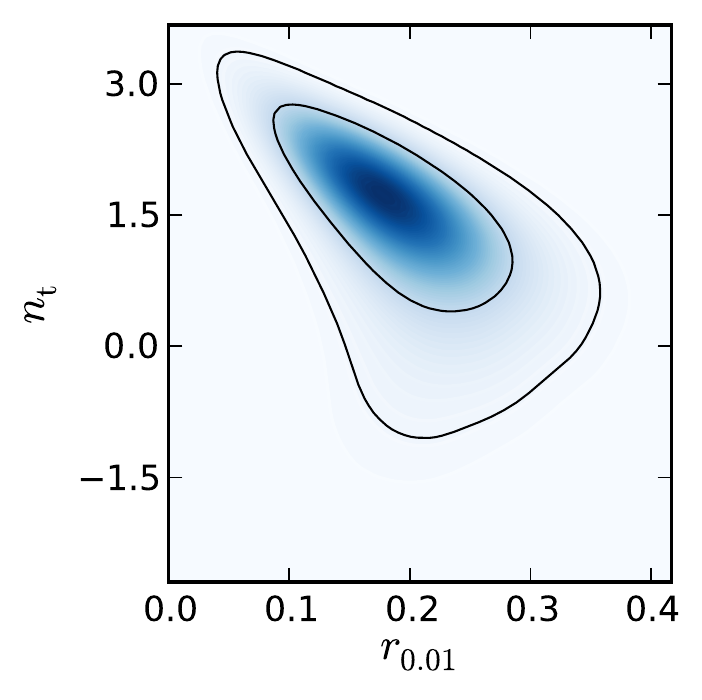}&\includegraphics[width=6.cm]{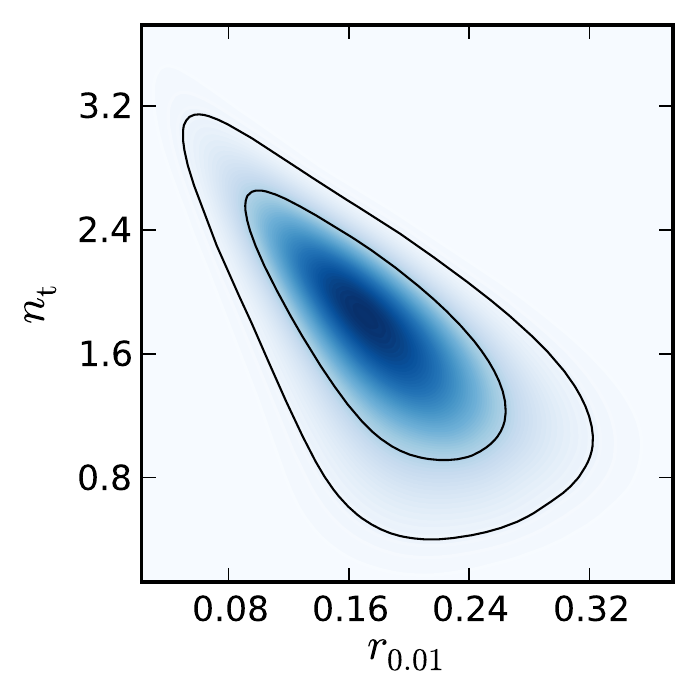}\\
\includegraphics[width=6cm]{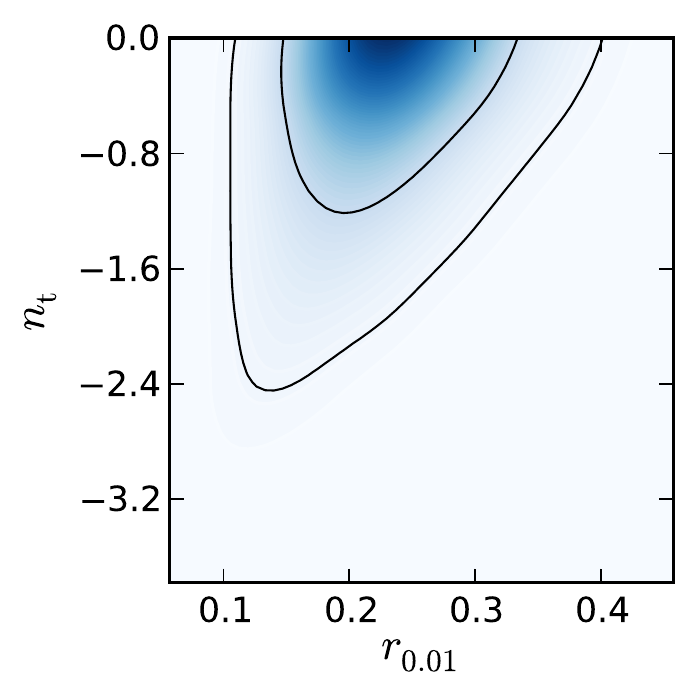}&\includegraphics[width=6.cm]{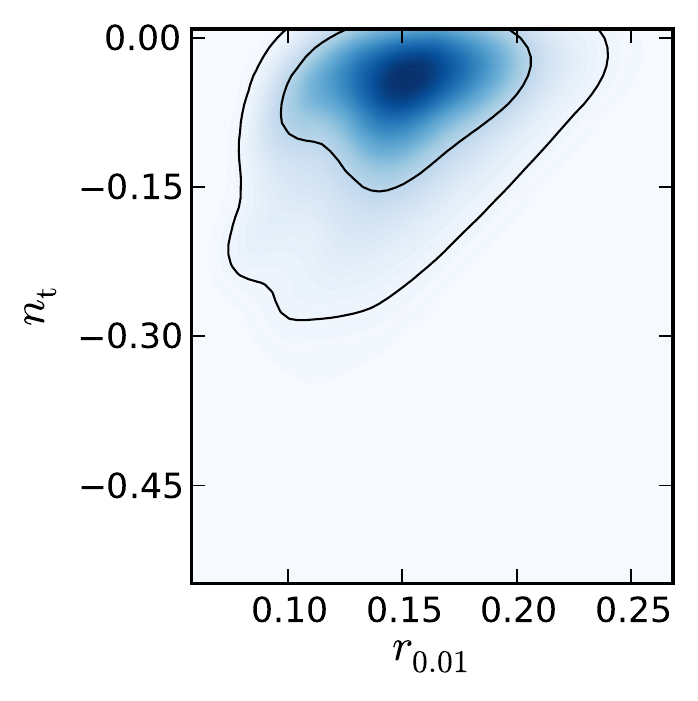}\\
\end{tabular}
 \caption{Constraints on the $n_T$ vs $r_{0.01}$ plane for the $4$ cases discussed in the
 analysis. No prior on $n_T$ (Top Left), No prior on $n_T$ but TT prior on $r_{0.002}$
 (Top Right), $n_T<0$ (Bottom Left), $n_T<0$ and TT prior on $r_{0.002}$ (Bottom Right)  
 }
\label{fig:bicep2}
\end{figure*}

\begin{table}[htb!]
\begin{tabular}{c||c|c}
\hline
\hline
Case&	$r_{0.01}$&				$n_T$\\
\hline
\hline
$n_T$ free&			$0.19\pm0.06$&		$1.36\pm0.83$\\
TT prior+$n_T$ free&			$0.18\pm0.05$&		$1.67\pm0.53$\\
$n_T<0$&			$0.22\pm0.06$&		$n_T>-0.76$\\
TT prior+$n_T<0$&			$0.15\pm0.03$&		$n_T>-0.09$\\
\hline
\hline
\end{tabular}
\caption{Constraints at $68 \%$ c.l. on $r_{0.01}$ and $n_T$ parameters for the cases
described in the text. A blue spectral index ($n_T>0$) is strongly suggested
when a TT prior of $r_{0.002} <0.11$ at $95 \%$ c.l. is included in the analysis.}
\end{table}

The results of our analysis are reported in Table I and Figure 1.
We consider four cases: $n_T$ free, $n_T$ free but with the TT prior,
$n_T$ assumed to be negative ($n_T<0$) and $n_T$ assumed to be negative
plus the TT prior.

We can derive the following conclusions:

\begin{itemize}

\item The BICEP2 data alone slightly prefers a positive spectral index. 
The case $n_T=0$ is consistent with the data in between two standard deviations.

\item When a TT prior of $r_{0.002}<0.11$ at $95 \%$ c.l.. is assumed, the BICEP2
data strongly prefers a blue spectral index with $n_T\le0$ excluded at more than
three standard deviations.

\item If we restrict the analysis to negative $n_T$ we obtain a lower limit
of $n_T>-0.76$ at $68 \%$ c.l. ($n_T>-0.09$ in case of the TT prior).
\end{itemize}

\section{Conclusions}\label{Conclusions}
In this brief note we have presented new constraints on the spectral index 
$n_T$ of tensor fluctuations from the recent data obtained by the BICEP2 experiment.
We found that the BICEP2 data alone slightly prefers a positive, "blue", spectral index with
$n_T=1.36\pm0.83$ at $68 \%$ c.l.. However, when a TT prior on the tensor
amplitude coming from temperature anisotropy measurements is assumed we get
$n_T=1.67\pm0.53$ at $68 \%$ c.l., ruling out a scale invariant $n_T=0$ 
spectrum at more than three standard deviations. 
Considering only the possibility of a "red", $n_T<0$ spectral
index we obtain the lower limit $n_T > -0.76$ at $68 \%$ c.l.
($n_T>-0.09$ when a TT prior is included).

These results are at odds with current upper limits on the tensor spectral
index coming from observations of pulsar timing, Big Bang Nucleosynthesis,
and from direct upper limits from the LIGO experiment (see e.g. \cite{upperlimitsnt}).
Considering $r_{0.002}=0.2$ and using the method adopted in \cite{upperlimitsnt} we 
found the current upper limits on $n_T$: $n_T \le 0.81$, $n_T \le 0.29$
and $n_T \le 0.15$ at $68 \%$ c.l. from pulsar timing, LIGO and 
BBN respectively.  The LIGO and BBN limits are in strong tension with
the BICEP2+CMB value. Therefore a positive spectral index does not provide 
an acceptable solution
to the tension between the BICEP2 data and current upper limits on $r$ from temperature
anisotropies. This indicates either the need of including extra parameters
(as the running of the scalar spectral index \cite{bicep2} or extra neutrino
species \cite{giusarma2014}) to relax current bounds on $r_{0.002}$ from temperature
anisotropies or the presence of unresolved 
systematics in current CMB data.

During the submission of this paper other works appeared discussing
the possibility of a BGW from BICEP2 (see \cite{bgw}) but without presenting 
numerical constraints on $n_T$ and an independent analysis of the BICEP2 data.
We also like to point out the discussion on the \texttt{cosmocoffee.info} 
website where results similar to ours have been presented by Antony Lewis.

\section*{Acknowledgments}

We like to thank Antony Lewis for the use of the numerical codes 
COSMOMC and CAMB.


\end{document}